\documentclass[letterpaper, 10 pt, conference]{ieeeconf}  
\IEEEoverridecommandlockouts                              
\usepackage{graphics} % for pdf, bitmapped graphics files
\usepackage{graphicx}
\usepackage{epsfig} % for postscript graphics files
\usepackage{times} % assumes new font selection scheme installed
\usepackage{balance}
\usepackage{color}
\usepackage{epstopdf}

\usepackage{amsmath}
\usepackage{amssymb}

\usepackage{booktabs}
\usepackage{mathrsfs}
\usepackage{theorem}
\usepackage{algorithm}
\usepackage{algorithmic}
\usepackage{tabularx}
\usepackage{indentfirst}

\newtheorem{proposition}{Proposition}
\newtheorem{theorem}{Theorem}

\newtheorem{remark}{Remark}

\newtheorem{assumption}{Assumption}

\newcommand{\Proof}{\noindent \textit{Proof.}$\;\;$}
\newcommand{\tr}{\intercal} % transpose

\title{\LARGE \bf Distributed State Estimation for AC Power Systems \\ using Gauss-Newton ALADIN
}

\author{Xu Du, Alexander Engelmann, Yuning Jiang, Timm Faulwasser and Boris Houska  % <-this % stops a space
\thanks{DX, YJ and BH are supported by ShanghaiTech University, Grant-Nr. F-0203-14-012.
	This work received funding from the European Union’s Horizon 2020
	research and innovation program under grant agreement No. 730936. TF acknowledges further
	support from the Baden-Württemberg Stiftung under the Elite Programme for Postdocs.}% <-this % stops a space
\thanks{XD, YJ and BH are with the School of Information Science and Technology, ShanghaiTech University, Shanghai, China
{\tt $\{$duxu, jiangyn, borish$\}$@shanghaitech.edu.cn }}%
\thanks{AE and TF are with the Institute for Automation and Applied Informatics, Karlsruhe Institue of Technology, Eggenstein-Leopoldshafen, Germany
{\tt alexander.engelmann@kit.edu timm.faulwasser@ieee.org}}%
}

\begin{document}

\maketitle
\thispagestyle{empty}
\pagestyle{empty}

%%%%%%%%%%%%%%%%%%%%%%%%%%%%%%%%%%%%%%%%%%%%%%%%%%%%
\begin{abstract}
This paper proposes a structure exploiting algorithm for solving  non-convex power system state estimation problems in  distributed fashion. Because the power flow equations in large electrical grid networks are non-convex equality constraints, we develop a tailored state estimator based on Augmented Lagrangian Alternating Direction Inexact Newton (ALADIN) method, which can handle the nonlinearities efficiently. Here, our focus is on using Gauss-Newton Hessian approximations within ALADIN in order to arrive at at an efficient (computationally and communicationally) variant of ALADIN for network maximum likelihood estimation problems. Analyzing an IEEE 30-Bus system we illustrate how the proposed algorithm can be used to solve highly non-trivial network state estimation problems. We also compare the method with existing distributed parameter estimation codes in order to illustrate its performance.
\end{abstract}

%%%%%%%%%%%%%%%%%%%%%%%%%%%%%%%%%%%%%%%%%%%%%%%%%%%%
\section{Introduction}
State estimation is of increasing importance in modern electricity transmission and distribution systems.
Due to the integration of renewable energy systems, effective grid operation often requires detailed knowledge of the system state.
High-accuracy measurement devices  are usually costly. Hence it is relevant to consider all available information and also cost-effective  (including possibly inaccurate) measurement devices for determining the power system's state.\footnote{We remark that in power systems the notion of \emph{state} variables differs slightly from control. Hence here we refer to a set of variables defined as the solution of a stationary nonlinear system of equations.}
A standard method to solve the arising Power System State Estimation (PSSE) problem is via weighted nonlinear least squares \cite{Schweppe1970,Abur2004,Monticelli2000}.

Centralized formulations of AC PSSE---i.e. considering the full AC power flow equations---have a long history and can be traced back to \cite{Schweppe1970b}.
AC PSSE is in general hard to solve as it is usually formulated as nonlinear least squares problem yielding a large-scale non-convex optimization problem.
Different formulations including polar vs. rectangular coordinates and algorithms with different Jacobian approximations, including exact Jacobian \cite{Schweppe1970}, $p$-$q$ decoupled Jacobian \cite{Horisberger1976}, and Gauss-Newton approximation \cite{Monticelli2000}, have been considered for the AC PSSE problem.

The non-convexity of the power flow equations makes large-scale PSSE problems difficult to solve. Hence several authors have considered convex formulations and relaxations of this problem. This includes DC approximations \cite{Schweppe1970a,Monticelli1999} and more recently SDP relaxations \cite{Zhang2016,Madani2016}.
However, as voltage and reactive power are often needed (especially in context of renewables) the practical usefulness of DC approximations is in general limited.

As power systems are large scale and as limited information exchange is desirable, distributed approaches have also been considered for AC and DC state estimation \cite{Schweppe1970b,Bin1994,Kekatos2013}.\footnote{Sometimes \textit{distributed} approaches are also called \textit{hierarchical} depending on the amount of central coordination \cite{Gomez-Exposito2011}.}
The DC case is considered in \cite{Schweppe1970a,Xie2012,Tai2013}.
In a distributed setting, AC PSSE is even harder to solve than in the centralized setting. The reason being that so far there are only a few algorithms for general distributed non-convex optimization  \cite{Boyd2011,Houska2016}.
Classical methods for distributed AC state estimation often exploit the sparsity pattern of the Jacobian of the measurement equations. % for developing distributed estimation schemes.
These works started already in the 1970s with the seminal paper \cite{Schweppe1970b}, continued with \cite{Bin1994,Iwamoto1989}; and can also be found today \cite{Korres2011}.
Recently \cite{Minot2016} proposed a distributed Gauss-Newton approach using matrix splitting techniques with promising results. 
Exploiting sparsity one obtains distributed methods where a coordinator typically solves preferably simple coordination problems.

A second and more recent line of research applies distributed optimization techniques coming from convex optimization to AC PSSE.
These approaches include the auxiliary problem principle \cite{Ebrahimian2000} and the popular Alternating Direction of Multipliers Method (ADMM) \cite{Kekatos2013}. 
An algorithm based on gossiping techniques can be found in \cite{Xie2012}.
These methods usually have an advantage over splitting techniques---they are typically decentralized, i.e. they avoid central coordination and communicate based on neighborhood information only.
However, despite working well for many cases, these methods usually have limited convergence guarantees for AC PSSE.
A recent overview on distributed AC state estimation can be found in \cite{Gomez-Exposito2011}.

In this paper, we follow a different route  tailoring the Augmented Lagrangian Alternating Direction Inexact Newton (ALADIN) method \cite{Houska2016} to AC PSSE problems. After introducing the problem at hand in Section~\ref{sec:ProbForm}, we explain how to exploit its distributed structure in Section~\ref{sec::Distributed}. Section~\ref{sec::ALADIN} introduces the main algorithmic contribution of this paper; i.e. we construct a variant of ALADIN based on generalized Gauss-Newton Hessian approximation. Such Gauss-Newton approximation based approaches have been analyzed exhaustively for unconstrained nonlinear least-squares problems \cite[Chapter 10.3]{Nocedal2006} as well as in the context of centralized parameter estimation of constrained problem as analyzed in~\cite{Bock1983,Bock2003} and~\cite{Schittkowski2016}. However, one interesting contribution of Section~\ref{sec::ALADIN} is analyzing such Hessian approximations in the context of distributed least squares estimation with the ALADIN framework proving a local convergence proof. The main contribution of this paper is presented in Section~\ref{sec::results}, where we not only illustrate the performance of the proposed algorithm on the IEEE 30-bus system in comparison to the widely used ADMM algorithm \cite{Boyd2011}, but also elaborate on the communication effectiveness of the proposed scheme.

\section{Power System State Estimation}
\label{sec:ProbForm}

This section briefly reviews the main physical relations in electrical grids, including power flow equations, and introduces the AC PSSE problem.

\subsection{Grid Model}
This paper considers a power system  $(\mathcal{N}, \mathcal{L},Y)$ consisting of a set of buses $\mathcal{N}=\{1,\dots,N\}$, a set of transmission lines $\mathcal{L}\subseteq \mathcal{N}\times \mathcal{N}$, and a sparse, complex-valued admittance matrix $Y=G+jB \in \mathbb{C}^{N\times N}$ with  $j=\sqrt{-1}$ . The admittance matrix is defined by 
\[
Y_{k,l}=\left\{
\begin{array}{ll}
\sum\limits_{l \neq k} \left( g_{k,l}+jb_{k,l} \right) & \text{if} \; k=l, \\  [0.25cm]
- \left( g_{k,l}+j b_{k,l} \right) & \text{if} \; k \neq l,
\end{array}
\right.
\]
where $g_{k,l}$ denotes the line conductance and $b_{k,l}$ denotes the line susceptance for all  transmission lines $(k,l) \in \mathcal L$. 
For $(k,l) \notin \mathcal L$, we have $g_{k,l} = b_{k,l} =0$.

To each node in the grid, we assign a state as
\begin{equation} \label{eq:nodState}
x_k^\top=(\theta_k \quad  v_k \quad p_k \quad q_k)^\top \in \mathbb{R}^4,
\end{equation}
where $\theta_i$ is the voltage angle,  $v_i$ is the voltage magnitude and $p_i,q_i$ are the net active and reactive power at node $i \in \mathcal{N}$.
The state of the grid is then defined as  $x^\top=(x_1,\dots,x_N)^\top \in \mathbb{R}^{4N}$.
The grid physics are described by the the power flow equations in polar form as 
\begin{subequations} \label{eq:PFeq}
\begin{align} 
0=p_k - v_k\sum_{l\in\mathcal{N}} v_l(G_{k,l}\cos(\theta_{k,l})+B_{k,l}\sin(\theta_{k,l})), \\
0 =q_k- v_k\sum_{l\in\mathcal{N}} v_l(G_{k,l}\sin(\theta_{k,l})-B_{k,l}\cos(\theta_{k,l})),
\end{align}
\end{subequations}
for all nodes $k\in \mathcal{N}$ with $\theta_{k,l}= \theta_k-\theta_l$, cf. \cite{Abur2004}. Note that $G_{k,l}$ and $B_{k,l}$ refer to the real and imaginary pats of the entries of the admittance matrix $Y$.

\subsection{Measurement Functions}
PSSE aims at determining the steady state, $x$, of the grid using a given set of measurements.
In general one considers two types of measurements: firstly one can directly measure the system states $x_k$ (or parts thereof) at the nodes. And secondly, one can attempt to measure the power flows at the transmission lines, which depend on the state of the grid at neighboring nodes. In order to arrive at a model that allows us to take the second type of measurements into account, we introduce so-called measurement functions, which relate the nodal states to the power flows over the transmission lines.
These functions are given by
\begin{equation*}
\label{eq::measure_funcs}
\begin{split} 
f_{p}(x_k,x_l) \;\;=&\;\; v_k[v_kg_{k,l}-v_lg_{k,l}\cos(\theta_{k,l})]\\
&\;\; -v_k[v_lb_{k,l}\sin(\theta_{k,l})  ]\;,\\[0.16cm]
f_{q}(x_k,x_l) \;\;=&\;\;-v_k[v_k b_{k,l}-v_lb_{k,l}\cos(\theta_{k,l})]\\
&\;\;+v_k[v_lg_{k,l}\sin(\theta_{k,l})]\;,\\[0.16cm]
f_{i}(x_k,x_l)\;\;=&\;\;\frac{f_{pt}(x_k,x_l)^2+f_{qt}(x_k,x_l)^2}{v_k^2},
\end{split}
\end{equation*}
where $f_{p},f_{q},f_{i}:\mathbb{R}^4\times\mathbb{R}^4\rightarrow \mathbb{R}$ denote the active power, respectively, reactive power and the current in the transmission line $(k,l) \in \mathcal{L}$. The complete vector-valued measurement function $\mathcal{F}_{\mathcal{N},\mathcal{L}}:\mathbb{R}^{4|\mathcal{N}|}\rightarrow \mathbb{R}^{4|\mathcal{N}|+3|\mathcal{L}|}$ is then given by
\begin{equation} \label{weight}
\mathcal{F}_{\mathcal{N},\mathcal{L}}(x):=
\begin{pmatrix}
\Sigma_k^\frac{1}{2} (x_k - \hat x_k)_{k \in \mathcal{N}}\\
W_{k,l}^\frac{1}{2} \left(f(x_k,x_l) - \hat w_{k,l}\right)_{(k,l)\in \mathcal{L}}
\end{pmatrix}
\end{equation}
where  $f=(f_{p},\;f_{q},\;f_{i})^\top$ is used. Moreover, we use the shorthand $\hat{w}_{k,l} = (\hat{p}_{k,l},\;\hat{q}_{k,l},\;\hat{i}^2_{k,l})^{\top}$ %, $\hat{p}_{k,l}$
to collect the actual measurements of the active and reactive power,
$\hat{p}_{k,l}$ and $\hat{q}_{k,l}$, as well as the measurement $\hat{i}_{k,l}$ of the current at the transmission line $(k,l)\in \mathcal{L}$. In this context, it is assumed that approximations of the inverse of variance matrices of the associated measurement errors, $\Sigma_{k}\in\mathbb{S}^{4}$ and $W_{k,l}\in\mathbb{S}^{3}$, are given by positive semi-definite matrices~\cite{Bock1983}.

\subsection{ Maximum Likelihood State Estimation}
The above model is used to formulate the AC PSSE problem of interest as the following nonlinear least-squares optimization problem
\begin{align}
\begin{aligned}
\min_{x}& \quad \|\mathcal{F}_{\mathcal{N},\mathcal{L}}(x)\|_2^2 \\
&\text{s.t. }  \eqref{eq:PFeq} \;\;\text{for all }k\in\mathcal{N}.
\label{eq:state_estimation}
\end{aligned}
\end{align}
Here, the underlying assumption is that the measurement errors have Gaussian probability distributions. This way~\eqref{eq:state_estimation} can be interpreted as a maximum likelihood parameter estimation problem~\cite{Bock2003,Houska2013}, recalling that the inverse variance matrices of the measurement errors, $\Sigma_{k}\in\mathbb{S}^{4}$ and $W_{k,l}\in\mathbb{S}^{3}$, are assumed to be given.

\begin{remark}
Although the theoretical properties of nonlinear least-squares optimization problems are rather well-understood~\cite{Ljung1999}, Problem~\eqref{eq:state_estimation} is in general a large-scale non-convex optimization problem with non-convex objective and non-convex constraint set over the complete electrical grid. As it turns out, AC PSSE problems are rather callenging to solve in practice. In particular, there might be multiple local minima and numerical algorithms might converge to one or the other minimum depending on the initialization \cite{Bukhsh2013}. 
\end{remark}

\section{Distributed Least Squares Estimation}
\label{sec::Distributed}
This section outlines the main strategy for breaking large-scale electrical grid networks into smaller sub-regions, thereby revealing the distributed structure of AC PSSE problems.% that will later be exploited by our numerical algorithms.

\subsection{Problem Decomposition}

\begin{figure}[htbp!]
	\centering
	\includegraphics[width=\linewidth]{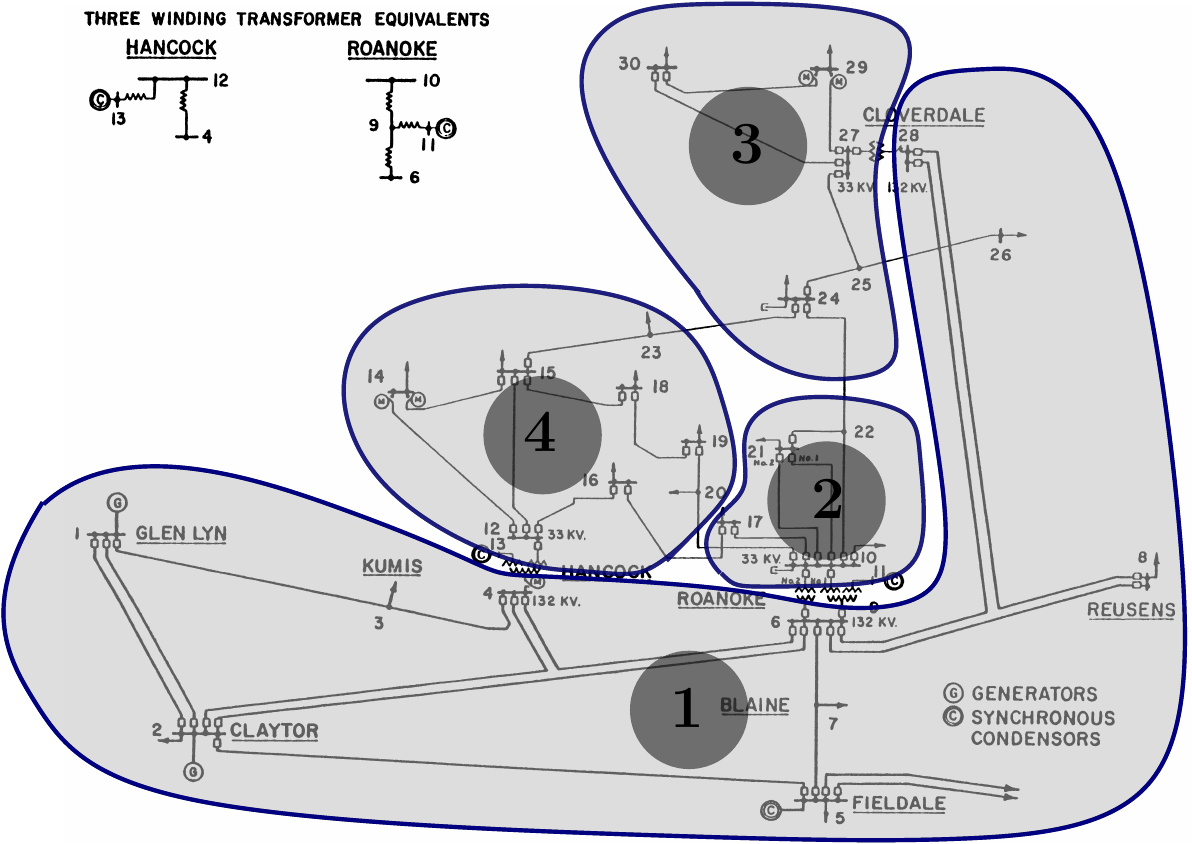}
	\caption{IEEE 30-bus system with partitioning.}
	\label{fig:30_bus}
\end{figure}

In order to solve \eqref{eq:state_estimation} in distributed fashion, we  reformulate \eqref{eq:state_estimation} in affinely-coupled separable from~\cite{Houska2016}. To this end, we recall the partitioning method from \cite{Engelmann18a}, for alternative partitioning schemes see \cite{Erseghe2015,Gomez-Exposito2011,Kekatos2013}. Figure~\ref{fig:30_bus} depicts the whole IEEE 30-Bus network as well as the partitioning strategy used throughout this paper.

\begin{figure}
	\centering
	\includegraphics[width=0.9\linewidth]{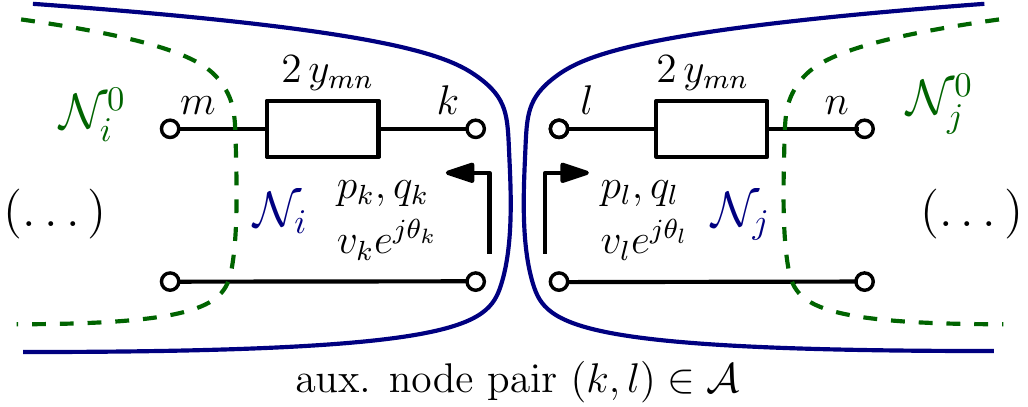}
	\caption{Decoupling of regions via auxiliary nodes.}
	\label{fig:splitting}
\end{figure}

We first divide the bus set $\mathcal{N}$ into several node sets $\mathcal{N}_i^0$, one for each region $\mathcal{R}=\{1,\dots R\}$ such that $\underset{i\in\mathcal{R}}{\cup}\mathcal{N}^0_i=\mathcal{N}$ and $\mathcal{N}^0_i\cap \mathcal{N}^0_j=\emptyset$ for all $i,j\in \mathcal{R}$ with $i \neq j$.
At each transmission line connecting two adjacent regions, i.e. all $(m,n)\in \mathcal{L}$ with $m\in \mathcal{N}_i$ and $n\in \mathcal{N}_j$, $i\neq j$, we introduce an auxiliary bus pair $(k,l)$ and we collect all auxiliary bus pairs in set $\mathcal{A}$.
The set of auxiliary buses of region $i \in \mathcal{R}$ are denoted as $\mathcal{A}_i$.
Finally, we combine all auxiliary nodes and original nodes belonging to one region in  combined node sets  $\mathcal{N}_i=\mathcal{N}_i^0 \cup \mathcal{A}_i$.
The line set connecting original nodes with each other and all auxiliary nodes for region $i\in \mathcal{R}$ is denoted by $\mathcal{L}_i$.
We assume a decomposition in the middle of each transmission line connecting two regions. This leads to new line admittances $y_{m,k}=2y_{m,n}$ and $y_{n,l}=2y_{m,n}$, respectively.
The partitioning strategy is graphically illustrated in Figure~\ref{fig:splitting} and Figure~\ref{fig:30_bus}.
In order to resemble the original physical properties of the grid model, we introduce the consensus constraints\footnote{Note that there exist different variants of coupling. Here we only couple voltages as we observed best performance in this case. However additional active/reactive power coupling is also possible \cite{Engelmann2019}, cf. \cite{Gomez-Exposito2011} for an overview of different coupling methods.}
\begin{align} 
\label{eq:consensus}
\theta_{k}\,=\,\theta_{l},\quad v_{k}\,=\,v_{l}, \text{ for all } (k,l)\in \mathcal{A},
\end{align}
a particular class of affine equality constraints.

\subsection{Distributed Formulation of the AC PSSE Problem}
This section reformulates \eqref{eq:state_estimation} in affinely coupled separable form, as required in the context of distributed optimization~\cite{Houska2016}. To this end, we introduce state vectors $z_i$ for all regions $i\in \mathcal{R}$ such that $z_i^\top=(\theta_i \quad  v_i \quad p_i \quad q_i)_{i\in \mathcal{N}_i}\in \mathbb{R}^{4|\mathcal{N}_i|}$.
Summarizing the measurement equations for all nodes $\mathcal{N}_i$ and transmission lines $(k,l)\in \mathcal{L}_i$ in each region $i\in \mathcal{R}$, i.e. $\mathcal{F}_{\mathcal{N}_i,\mathcal{L}_i}$ yields
\begin{subequations}\label{eq::reformulation}
\begin{align}
\min_{z} \quad\sum_{i\in\mathcal{R}} \|\mathcal{F}&_{\mathcal{N}_i,\mathcal{L}_i}(z_i)\|_2^2\\ \text{subject to}\,\quad\sum_{i\in\mathcal{R}} A_i z_i &= 0 \quad \mid \lambda \label{eq:consConstr}\\ 
\quad  \mathcal{H}_i(z_i)&=0 \;\;\text{for all } i\in \mathcal{R}, \notag
\end{align}
\end{subequations}
with $z^\top = (z_1^\top,\dots,z_R^\top)\in \mathbb{R}^{4|\mathcal{N}|}$, \eqref{eq:consConstr} contains equations \eqref{eq:consensus} by appropriate choice of coupling matrices $A_i\in \mathbb{R}^{2|\mathcal{A}|\times4|\mathcal N|}$ and $\lambda$ denotes Lagrange multipliers assigned to \eqref{eq:consConstr}.
Subsequently we denote $\mathcal{F}_{\mathcal{N}_i,\mathcal{L}_i}$ as $\mathcal{F}_i$ for simplicity.
The equality constraint $\mathcal{H}_i$ collects the power flow equations $\eqref{eq:PFeq}$ for all $i \in \mathcal{R}$.

\section{Distributed Optimization Algorithm}
\label{sec::ALADIN}

This section introduces a Gauss-Newton Hessian approximation based variant of the distributed optimization algorithm ALADIN~\cite{Houska2016}, which is tailored to AC PSSE problems in nonlinear least-squares form.

\begin{algorithm}[htbp!]
	\caption{Gauss-Newton ALADIN }
	\textbf{Initialization:} Initial guess $(z,\lambda)$, choose $\rho,\epsilon>0$. \\
	\textbf{Repeat:}
	\begin{subequations}
	\begin{enumerate}
		\item \textit{Parallelizable Step:} \label{step:parStep}
		Solve for each $i \in \mathcal{R}$
		\begin{align}
		\label{eq::deNLP}
		\begin{split}
		\underset{y_i}{\min}&\;\;\; \left\| \mathcal{F}_i(y_i) \right\|_2^2 + \lambda^\top A_i y_i +  \frac{\rho}{2}\left\|y_i-z_i\right\|_{2}^2\\
		\text{s.t.}&\quad\mathcal{H}_i(y_i) = 0\quad \mid \kappa_i^d
		\end{split}
		\end{align}
		and compute 
		\begin{equation}
		\label{eq::senEval}
		b_i = \mathcal{F}_i(y_i),\; B_i = \nabla \mathcal{F}_i(y_i)^\top,\;C_i = \nabla \mathcal{H}_i(y_i)^\top
		\end{equation}
		in parallel.
		
		\item \textit{Termination Criterion:} \label{step:terStep}
		Terminate if 
		\begin{equation}
		\label{eq::stop}
		\left\|\sum_{i\in \mathcal{R}}A_iy_i\right\|\leq \epsilon \text{ and } \left\| y_i - z_i \right \|_\infty\leq \epsilon\;.
		\end{equation}

		\item \textit{Consensus Step:} \label{step:conStep}
		Solve the coupled QP problem  
		\begin{equation}
		\label{eq::cQP}	
		\begin{split}
		\underset{\Delta y}{\min}&\quad \sum_{i\in \mathcal{R}}\left\|B_i\Delta y_i\right\|_2^2 + 2\Delta y_i^\top B_i b_i \\ 
		\text{s.t.}\;&\quad \sum_{i\in \mathcal{R}}A_i(y_i+\Delta y_i) = 0 \quad \mid  \lambda^\mathrm{QP},\\
		&\quad\; C_i \Delta y_i = 0 \quad \mid\kappa_i^\mathrm{QP}\quad i\in \mathcal{R}.
		\end{split}
		\end{equation}
		and update $z^{+} \leftarrow y + \Delta y$, $\lambda^+ \leftarrow \lambda^\mathrm{QP}$.
	\end{enumerate}
	\label{alg:ALADIN}
	\end{subequations}
\end{algorithm}

\subsection{Main Algorithmic Steps}
Algorithm~\ref{alg:ALADIN} outlines a variant of ALADIN for solving~\eqref{eq::reformulation}. Similar to the traditional ALADIN procedure, there are two main steps: A parallelizable Step~\ref{step:parStep}) and a consensus Step~\ref{step:conStep}). In Step~\ref{step:parStep}), decoupled NLPs \eqref{eq::deNLP} are solved followed by a sensitivity evaluation \eqref{eq::senEval}---both in parallel.
Note that due to the Gauss-Newton Hessian approximation, we only need to compute first-order derivatives. This way the computational burden and communication overhead is reduced significantly compared with standard ALADIN.

\subsection{Local Convergence Analysis}
Let $(z^*,\lambda^*,\kappa^*)$ denote a primal-dual locally optimal solution of~\eqref{eq::reformulation}, where $\lambda^\star$ denotes the multiplier of the linear coupling constraints and $\kappa^\star$ the multiplier that is associated with the power-flow equations. In the following, we introduce the following regularity assumption on the physical power flow equations.

\begin{assumption}
\label{ass::LICQ}
The Jacobian of the power flow equations~\eqref{eq:PFeq} with respect to all states $x$ of the network at $z^\star$ has full row-rank.
\end{assumption}

Notice that a detailed discussion of mathematical conditions under which the linear inpendendence constraint qualification (LICQ) condition in Assumption~\ref{ass::LICQ} is satisfied for power flow networks can be found in~\cite{Hauswirth18a}, where it is also discussed why this assumption is essentially satisfied for all power-flow networks of practical interest.

\begin{proposition}
\label{prop::LICQ}
If Assumption~\ref{ass::LICQ} holds, then the LICQ condition for the decoupled NLPs~\eqref{eq::deNLP} as well as the coupled QP~\eqref{eq::cQP} is satisfied, that is, the matrix $[A^\tr\;\; C^\tr]^\tr$ has full row rank.
\end{proposition}

\Proof
The proof of the proposition follows from the fact that the Jacobian of the consensus constraint has---by definition---full rank, as this constraint enforces linear coupling between neighboring regions. Because the power-flow equations are local (decoupled) in the reformulated problem~\eqref{eq::reformulation}, they satisfy the decoupled LICQ conditions (since Assumption~\ref{ass::LICQ} holds), and, additionally, cannot possibly be redundant to the coupling constraints. 
%This implies the statement of this proposition.
\hfill$\blacksquare$

\smallskip
\noindent
In order to further ensure that any local solution $(z^*,\lambda^*,\kappa^*)$ is a regular KKT point of~\eqref{eq::reformulation}, the following proposition is introduced.

\begin{proposition}
\label{prop::SOSC}
If the residuum $\sum_{i} \Vert \mathcal F_i(z_i^\star) \Vert_2^2$ in the optimal solution is sufficiently small and if the matrices $\Sigma_k$ are positive definite, then the second order sufficient optimality condition (SOSC) is satisfied for~\eqref{eq::reformulation} at $z^\star$ and the Gauss-Newton Hessian approximation,
$\nabla \mathcal{F}_i(z_i) \nabla \mathcal{F}_i(z_i)^\top\succ 0$ is positive definite in a local neighborhood of $z^\star$.
\end{proposition}

\Proof
The statement of the above proposition is well-known in the context of Gauss-Newton SQP methods and a formal proof can be found in~\cite{Bock1983}. We remark that the conditions therein are indeed satisfied if the matrices $\Sigma_k$ are positive definite, as this condition trivially ensures identifiability of all measured states.\hfill$\blacksquare$

\smallskip
\noindent
Note that the conditions of the above proposition are satisfied in practice if the model-data mismatch is \mbox{small---but} it can be violated otherwise.

\begin{theorem}
\label{thm:conv}
Let Assumption~\ref{ass::LICQ} be satisfied and let the residuum $\sum_{i} \Vert \mathcal F_i(z_i^\star) \Vert_2^2$ at the local minimizer be sufficiently small such that Propsition~\ref{prop::SOSC} is applicable. Then the iterates $(z,\lambda)$ locally converge to $(z^*,\lambda^*)$ achieving a locally linear convergence rate.
\end{theorem}

\Proof
Propositions~\ref{prop::LICQ} and~\ref{prop::SOSC} ensure that minimizers of the decoupled NLPs~\eqref{eq::deNLP} are regular KKT points in a neighborhood of the optimal solution. Hence we can apply
Lemma~3 in~\cite{Houska2016} to show that the solution $(y,\kappa^d)$ of the decoupled NLP satisfies
\begin{equation}
\label{eq::local_Lipschitz}
\left\|
\left[
\begin{array}{c}
y-z^*\\
\kappa^d - \kappa^*
\end{array}
\right]
\right\|_2
\leq
\alpha\left\|
\left[
\begin{array}{c}
z-z^*\\
\lambda - \lambda^*
\end{array}
\right]
\right\|_2
\end{equation}
for a constant $\alpha<\infty$. Furthermore, in~\cite{Houska2016} it has been shown that the consensus step of ALADIN is locally equivalent to one SQP iteration. Thus, as we employ a Gauss-Newton Hessian approximation, we have
\[
\left\|
\left[
\begin{array}{c}
z^+-z^*\\
\lambda^+ - \lambda^*
\end{array}
\right]
\right\|_2\leq \gamma 
\left\|
\left[
\begin{array}{c}
y-z^*\\
\lambda-\lambda^*\\
\kappa^d - \kappa^*
\end{array}
\right]
\right\|_2\; .
\]
Next, recall that the linear convergence rate of Gauss-Newton methods is locally proportional to the least-squares residuum at the optimal solution. In other words, we have $\gamma = \mathbf{O}( \sum_{i} \Vert \mathcal F_i(z_i^\star) \Vert_2^2 )$, as proven in~\cite{Bock1983}. Thus, as long as the least-squares residuum is sufficiently small, it holds that
\[
\left\|
\left[
\begin{array}{c}
z^+-z^*\\
\lambda^+ - \lambda^*
\end{array}
\right]
\right\|_2\leq \gamma(\alpha+1) 
\left\|
\left[
\begin{array}{c}
z-z^*\\
\lambda-\lambda^*
\end{array}
\right]
\right\|_2
\]
with $\gamma(\alpha+1) < 1$. This finishes the proof. %corresponds to the statement of the theorem. 
\hfill$\blacksquare$

\smallskip
\noindent
Note that Theorem~\ref{thm:conv} establishes local convergence of Algorithm~1 only. Thus, if one has poor initial guesses for the state, the proposed method needs to be augmented by a globalization routine, as discussed in~\cite{Houska2016}. This is subject to future work.

\subsection{Communication Overhead}
Step~\ref{step:conStep} of Algorithm~\ref{alg:ALADIN} communicates between different regions. The forward communication collects matrices $B_i^\top B_i$, $C_i$ and vectors $B_ib_i$, $A_iy_i$ such that there are in total 
\[
\sum_{i\in\mathcal{R}}6|\mathcal{N}_i| +16|\mathcal{N}_i|^2 + 2|\mathcal{A}|
\]
floats that  need to be uploaded. The download phase after~\eqref{eq::cQP} is solved sends the dual update $\lambda^+$ and local direction $\Delta y_i$ to each region, which requires $2|\mathcal{A}|+4|\mathcal{N}_i|$ floats in total for each region $i\in \mathcal{R}$.

\section{Numerical Example}
\label{sec::results}

In this section, we illustrate the performance of Algorithm~\ref{alg:ALADIN} drawing upon the 30-bus system shown in Figure~\ref{fig:30_bus}.

\subsection{Implementation and Data}
The problem data is obtained from the  \texttt{MATPOWER} dataset~\cite{Zimmerman2011}, although in our case study
shunt elements are neglected. The system is partitioned into four regions $\mathcal{R}=\{1,2,3,4\}$ which are linked by $|\mathcal{A}|=8$
pairs of auxiliary nodes. We use nodal measurements and line measurements for all original nodes
$k\in \mathcal{N}^0$ and all lines connecting original nodes $(k,l)\in \mathcal{N}_i\setminus\mathcal{A}_i \times \mathcal{N}_i\setminus\mathcal{A}_i$ and all $i \in \mathcal{R}$.

The measurements in our case study have been obtained by running a realistic scenario simulation in \texttt{MATPOWER}. During this simulation, we have introduced an additional Gaussian white noise with zero mean and a relative error variance of $10^{-4}$ the states $\theta_k, p_k,q_k$. In addition,  the relative variance of the noise added to the voltage magnitude $v_k$ has been set to $10^{-5}$. Notice that such a white noise as been added for all $k\in \mathcal{N}^0$. The flows over transmission lines $p_{k,l}, q_{k,l}$ and $i_{k,l}$ are subject to a relative variance of $10^{-5}$, which is a standard value that is often used in the context of PSSE~\cite{Wood2013}. The considered associated weighting matrices of the least-squares objective are
\[ \Sigma_k= \operatorname{cov}(\hat x_k)^{-1}=\begin{bmatrix} 10^4&0&0&0\\0&10^5&0&0\\
0&0&10^4&0\\0&0&0&10^4 \end{bmatrix}\quad , \]
for all nodes $k\in \mathcal{N}^0$ and
\[W_{k,l}=\operatorname{cov}(\hat w_k)^{-1} = \begin{bmatrix} 10^4&0&0\\0&10^4&0\\
0&0&10^4 \end{bmatrix}\quad\]
for all  $(k,l)\in \mathcal{N}_i\setminus\mathcal{A}_i \times \mathcal{N}_i\setminus\mathcal{A}_i$. It can be checked numerically that this choice ensures that the local convergence conditions of Theorem~\ref{thm:conv} are indeed satisfied.

\subsection{Numerical Comparison of ADMM and ALADIN}
The implementation of Algorithm~1 relies on \texttt{Casadi-v3.4.5} with \texttt{IPOPT} and \texttt{MATLAB 2018a}. The tuning parameters in Algorithm~\ref{alg:ALADIN} are set to $\rho=10^{4}$ and $\epsilon= 10^{-4}$. Moroever, in order to assess the numerical performance of the proposed Gauss-Newton ALADIN algorithm, we compare our implementation with a standard implementation of ADMM, where the augmented Lagrangian parameter is set to $\rho^{\text{ADM}} = 10^{4}$, too. Note that ADMM does provide convergence guarantees for general non-convex problems. Indeed counter-examples where ADMM is divergent are given in~\cite{Houska2016}. However, it turns out that both ADMM and Gauss-Newton ALADIN converge for this particular PSSE case study.

\begin{figure}
	\centering
	\includegraphics[width=1.1\linewidth]{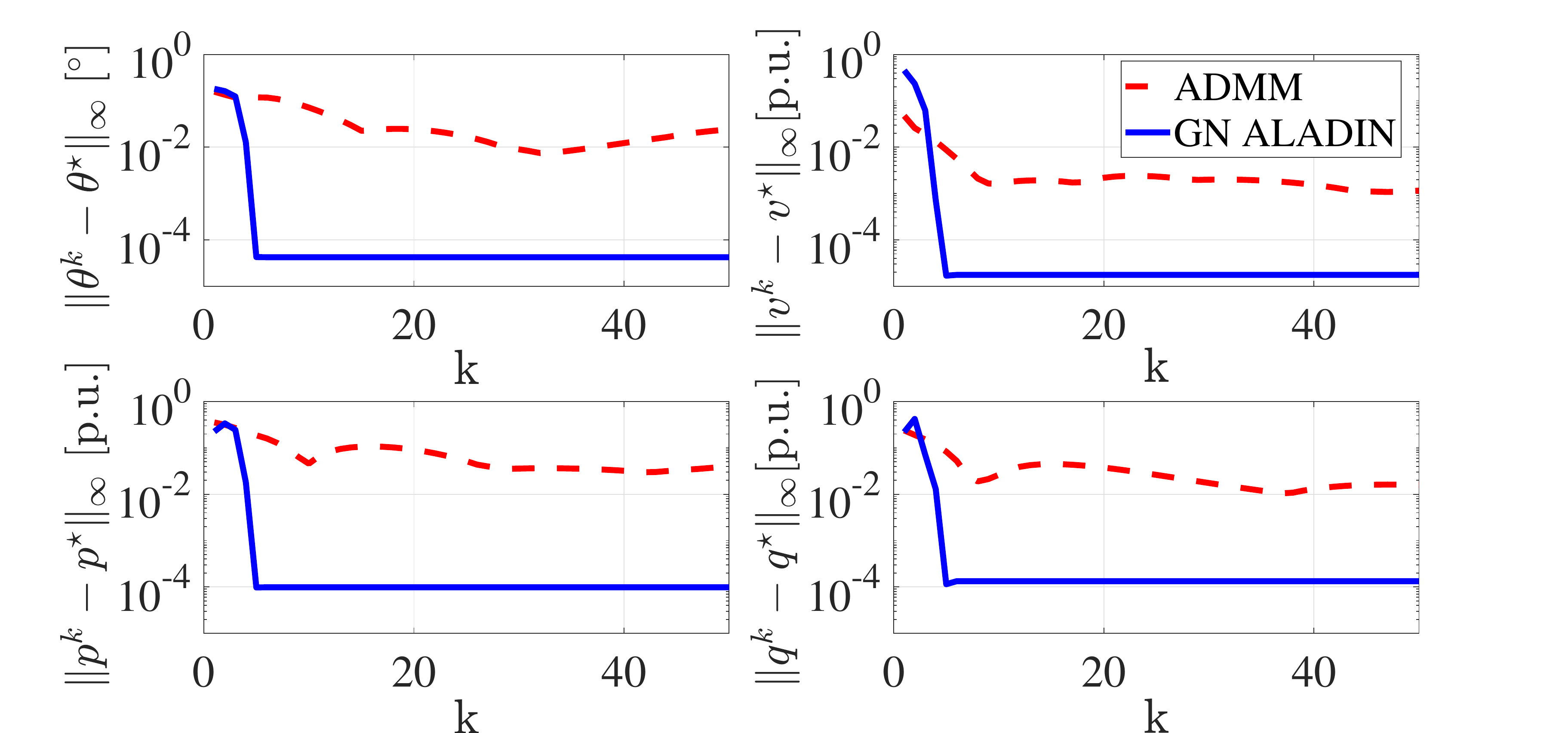}
	\caption{Convergence of states for the  IEEE 30-bus system. }
	\label{fig:variable}
\end{figure}
\begin{figure}
	\centering
	\includegraphics[trim={0em 0.5em 0 0},clip,width=0.7\linewidth]{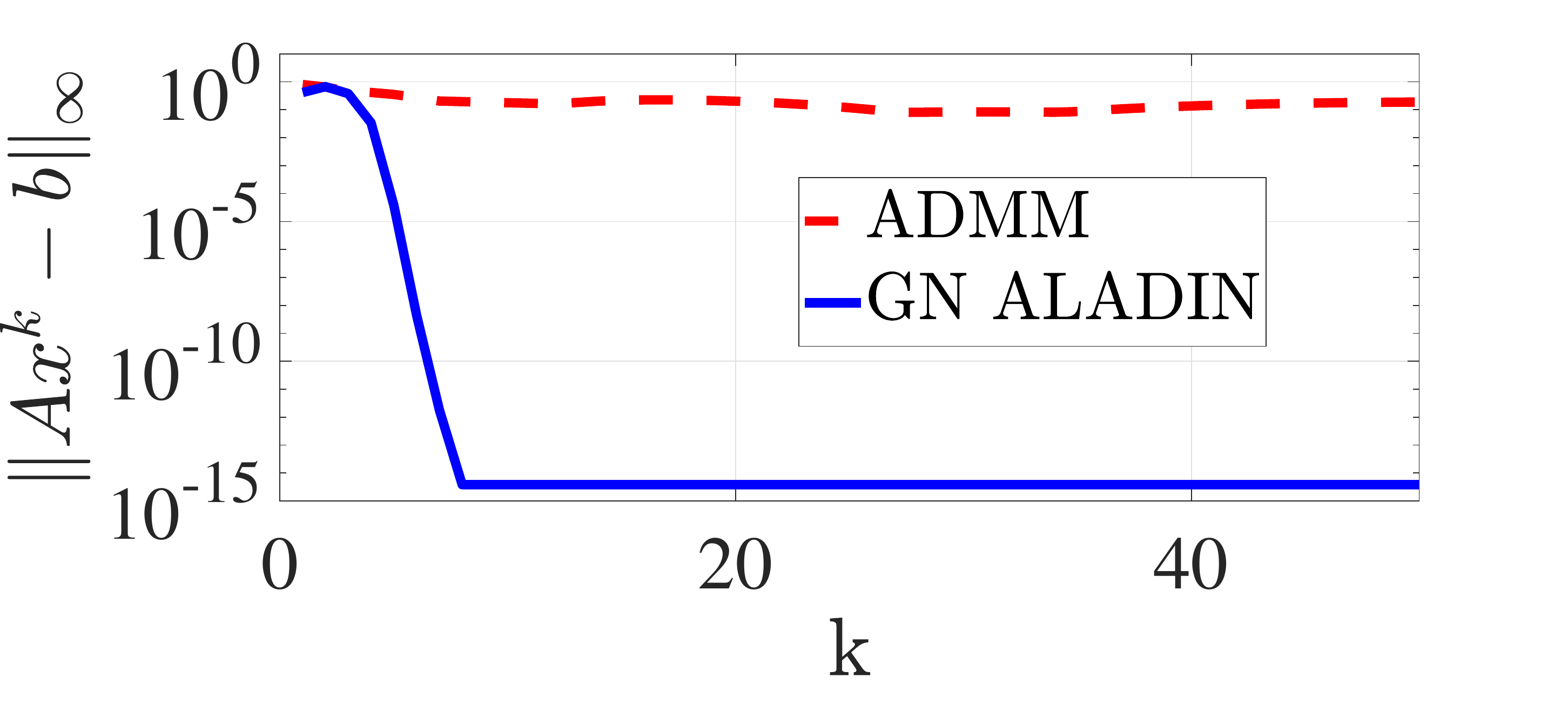}
	\caption{Consensus violation.}
	\label{fig:Ax-b}
\end{figure}

Figure \ref{fig:variable} shows the convergence of voltage angles, voltage magnitudes and active/reactive powers over the iteration index $k$.
One can observe that ALADIN converges at a fast linear rate while ADMM converges quite slower  to lower accuracy.
%\footnote{It is well-known that ADMM is rather scaling dependent~\cite{Boyd2011}. In our numerical implementation all variables have already been scaled before running the ADMM routine, but, of course, we cannot exclude that it is possible to further improve the performance of ADMM by developing more sophisticated scaling or pre-conditioning strategies. However, one of the key advantages of Algorithm~1 versus ADMM is that it works ``out of the box''; that is, there is no pre-conditioning or scaling needed, as Gauss-Newton methods are naturally invariant with respect to scaling~\cite{Bock1983,Houska2016}.}
Figure \ref{fig:Ax-b} shows the convergence of the corresponding consensus constraint violation $\|Ax^k-b\|_\infty$, which can be interpreted as the degree of matching of the voltage angles and magnitudes at auxiliary nodes according to \eqref{eq:consensus}.
The fast linear convergence of ALADIN can also here be witnessed; that is, at least for this PSSE problem, our numerical results indicate a much better convergence of ALADIN when compared to ADMM. However, as shown in \cite{Engelmann2019}, one should keep in mind that ALADIN has a higher per-step communication overhead and complexity compared to ADMM.
\begin{remark}[Effects of scaling on convergence]
It is well-known that ADMM is rather scaling dependent~\cite{Boyd2011}. In our numerical implementation all variables have already been scaled before running the ADMM routine, but, of course, we cannot exclude that it is possible to further improve the performance of ADMM by developing more sophisticated scaling or pre-conditioning strategies. However, one of the key advantages of Algorithm~1 versus ADMM is that it works ``out of the box''; that is, there is no pre-conditioning or scaling needed, as Gauss-Newton methods are naturally invariant with respect to scaling~\cite{Bock1983,Houska2016}.
\end{remark}

\subsection{A Posteriori Error Analysis}
As for Bayesian inference or maximum likelihood estimation problem, there arises also in PSSE the question what can be said about the quality of the a posteriori distribution of the parameter estimate. At this point, we rely on a mature body of literature on nonlinear parameter estimation theory as reviewed in~\cite{Ljung1999}. Therein, it has been proven that the inverse of the Fisher information matrix of a nonlinear least-squares parameter estimation problem is as a lower bound of the a-posteriori parameter estimation variance matrix. This relation is also known as \textit{Cram\'er-Rao bound}. We refer to~\cite{Ljung1999,Houska2013} for an in-depth discussion and further references. Note that the inverse Fisher information matrix of the state estimate of the $i$-th subregion is in our context given by
\[
\begin{bmatrix}
I \\[0.1cm] 0 \\[0.1cm] 0
\end{bmatrix}^\top
\begin{bmatrix} B_{\mathcal R}^\top B_{\mathcal R}  & C_{\mathcal R}^\top & A_{\mathcal R}^\top \\[0.1cm]
C_{\mathcal R}^\top & 0 & 0 \\[0.1cm]
A_{\mathcal R}^\top & 0 & 0
\end{bmatrix}^{-1} \begin{bmatrix}
I \\[0.1cm] 0 \\[0.1cm] 0
\end{bmatrix} \; ,
\]
where  $B_{\mathcal R} = \mathrm{diag}(B_i)_{i \in \mathcal R}$ denotes the derivative of $\mathcal F$ while $A_{\mathcal R}$ and $C_{\mathcal R}$ denote the associated constraint Jacobian matrices, all evaluated a-posteriori at the optimal solution. A detailed derivation of this expression for the inverse Fisher information matrix in the context of constrained Gauss-Newton methods can be found in~\cite{Bock2003}, see also~\cite{Houska2013}. The square-roots of selected diagonal elements of the above matrix relative to the nominal value of the associated parameter estimate can be found in Table~\ref{TABLE2}.
\renewcommand{\arraystretch}{1.5} 
\begin{table}[!htbp]  	
	\centering  
	\fontsize{7.5}{8}\selectfont 
	\caption{Relative a-posteriori standard deviations at selected nodes.} 
	\begin{tabular}{lllll}  
		\hline  
		\multicolumn{4}{c}{Relative Standard Deviation of the A-Posteriori Distribution}\cr%\cline{2-9}  
		Bus$\# \quad$ &$\; \; \theta \qquad \qquad$&$v \qquad \qquad$&$p \qquad \qquad$&$q$\cr  
		\hline  
		1&$*$     &$0.12\%$&$0.16\%$     &$2.65\%$\cr  
		8&$0.32\%$&$0.12\%$&$0.16\%$&$0.23\%$\cr  
		13&$0.65\%$&$0.17\%$&$0.05\%$&$0.79\%$\cr  
		20&$0.25\%$&$0.18\%$&$0.46\%$&$14.16\%$\cr  
		30&$0.33\%$&$0.23\%$&$0.28\%$&$2.95\%$\cr
		AVG&$0.49\%$&$0.17\%$&$0.98\%$&$2.75\%$\cr  	
		\hline  
	\end{tabular}  
	\label{TABLE2}
\end{table}  
As there are $30$ busses in total not all values are shown. However, the last line lists the average values for the standard deviation over the whole network. Since most of the relative errors are below $1 \%$, it can certainly be claimed that no over-fitting effects are visible in our PSSE case study. Nevertheless, some of the a-posteriori standard deviations of the reactive power estimates are around $14 \%$, which indicates that some of the states of this 30-bus power system are rather difficult to estimate from the measurement data. In fact, these results suggest that a more detailed analysis of the parameter estimation accuracy in PSSE, as well as the optimization of sensor locations in power grids are an interesting future research direction.

\section{Conclusion \& Outlook}
This work has introduced a distributed state estimation algorithm for non-convex AC PSSE
problems based on ALADIN and a Gauss-Newton Hessian approximation. A local convergence
condition for this algorithm has been given in Theorem~\ref{thm:conv}. Moreover, we have
illustrated the promising convergence behavior of Gauss-Newton compared to state-of-the-art
ADMM methods by analyzing a highly non-trivial IEEE $30$-bus power grid.

\bibliographystyle{plain}
\bibliography{paper}
\balance

\end{document}